 \documentstyle[12pt,multibox]{article}
 \def\a{\alpha}
 \def\b{\beta}
 \def\g{\gamma}
 \def\d{\delta}
 \def\e{\eta}
 \def\h{\eta}
 \def\l{\lambda}

 \def\m{\mu}
 \def\n{\nu}
 \def\r{\rho}
 \def\o{\omega}
 \def\s{\sigma}

 \def\x{\xi}
 \def\e{\varepsilon}

 \def\pa{\partial}

 \def\be{\begin{equation}}
 \def\ee{\end{equation}}
 \def\beq{\begin{eqnarray}}
 \def\eeq{\end{eqnarray}}

 \def\cl{{\cal L}}

 \newcommand{\bqn}{\begin{eqnarray}}\newcommand{\eqn}{\end{eqnarray}}
 \newtheorem{theorem}{Theorem}[subsection]

\begin{document}

 \begin{titlepage}
 \begin{flushright}
 DFPD/02/TH/25 \\
 ULB-TH-02/31 \\
 \end{flushright}
 \vskip 1.0cm

 \begin{centering}

 {\large {\bf Consistent deformations of dual formulations of 
linearized gravity: A no-go result}}

 \vspace{1cm}
 Xavier Bekaert$^{a}$,
 Nicolas Boulanger$^{b,}\footnote{``Chercheur F.R.I.A., Belgium''}$ and
 Marc Henneaux$^{b,c}$ \\
 \vspace{.7cm}
 {\small
 $^a$ Dipartimento di Fisica, Universit\`a degli Studi di Padova,
 Via F. Marzolo 8, I-35131 Padova, Italy \\
 \vspace{.2cm}
 $^b$ Physique Th\'eorique et Math\'ematique,
 Universit\'e Libre
 de Bruxelles,  C.P. 231, B-1050, Bruxelles, Belgium      \\
 \vspace{.2cm} $^c$ Centro de Estudios Cient\'{\i}ficos, Casilla
 1469, Valdivia, Chile }

 \vspace{.5cm}

 \end{centering}

 \begin{abstract}
 The consistent, local, smooth deformations of the dual formulation
 of linearized gravity involving a tensor field in the exotic
 representation of the Lorentz group with Young symmetry type
 $(D-3,1)$ (one column of length $D-3$ and one column of length
 $1$) are systematically investigated. The rigidity of the Abelian
 gauge algebra is first established. We next prove a no-go theorem
 for interactions involving at most two derivatives of the fields.
 \end{abstract}

 \vfill
 \end{titlepage}

 \section{Introduction}
 \setcounter{equation}{0} \setcounter{theorem}{0}
 \setcounter{lemma}{0} The electric-magnetic duality is one of the most
 fascinating symmetries of theoretical physics.  Recently
 \cite{Hull1}, dual formulations of linearized gravity \cite{Curt}
 have been
 systematically investigated with $M$-theory motivations in mind
 \cite{Hull2,Hull3} (see also \cite{CMR}).  These dual formulations
 involve tensor fields in ``exotic" representations of the Lorentz
 group characterized by a mixed Young symmetry type.  There exist
 in fact three different dual formulations of linearized gravity in
 generic spacetime dimension $D$. The first one is the familiar
 Pauli-Fierz description based on a symmetric tensor $h_{\m \n}$.
 The second one is obtained by dualizing on one index only and
 involves a tensor $T_{\l_1 \l_2 \cdots \l_{D-3} \m}$ with
 \begin{eqnarray}
 && T_{\l_1 \l_2 \cdots \l_{D-3} \m} = T_{[\l_1 \l_2 \cdots
 \l_{D-3}] \m}, \\
 &&T_{[\l_1 \l_2 \cdots \l_{D-3} \m]} = 0
 \end{eqnarray}
 where square brackets denote antisymmetrization with strength one.
 {}Finally, the third one is
 obtained by dualizing on both indices and is described by a tensor
 $C_{\l_1 \cdots \l_{D-3} \m_1 \cdots \m_{D-3}}$ with Young
 symmetry type $(D-3,D-3)$ (two columns with $D-3$ boxes). Although
 one can write equations of motion for this theory which are
 equivalent to the linearized Einstein equations, these do not seem
 to follow (when $D>4$) from a Lorentz-invariant action principle
 in which the only varied field is $C_{\l_1 \cdots \l_{D-3} \m_1
 \cdots \m_{D-3}}$. For this reason, we shall focus here on the
 dual theory based on $T_{\l_1 \l_2 \cdots \l_{D-3} \m}$.

 The purpose of this paper is to determine all the consistent,
 local, smooth interactions that this dual formulation admits.  It
 is well known that the only consistent (local, smooth) deformation
 of the Pauli-Fierz theory is - under quite general and reasonable
 assumptions - given by the Einstein theory (see \cite{others} and
 the more recent works \cite{Wald,BDGH} for systematic analyses).
 Because dualization is a non-local process, one does not expect
 the Einstein interaction vertex to have a local counterpart on the
 dual $T_{\l_1 \l_2 \cdots \l_{D-3} \m}$-side.  This does not a
 priori preclude the existence of other local interaction vertices,
 which would lead to exotic self-interactions of ``spin-2"
 particles. Our main -- and somewhat disappointing -- result is,
 however, that this is not the case.

 The first instance for which $T_{\l_1 \l_2 \cdots \l_{D-3} \m}$
 transforms in a true
 exotic representation of the Lorentz group
 occurs for $D=5$, where one has
 $$T_{[\a\b ]\g}  \simeq \hbox{
 \footnotesize \begin{picture}(38,15)(0,0)
 \multiframe(10,3)(10.5,0){2}(10,10){$\b$}{$\gamma$}
 \multiframe(10,-7.5)(10.5,0){1}(10,10) {$\a$}
 \end{picture}\normalsize}   . $$
 The action of this dual theory is given in \cite{Curt}(see also
 \cite{Aula,Lab1,Lab2}).  We shall explicitly investigate the
 $T_{[\a\b ]\g}$-case in this paper and comment on general gauge fields
 $T_{\l_1 \l_2 \cdots \l_{D-3} \m}$ at the end.

 Our precise result is that the free field dual theory based
 on $T_{\l_1 \l_2 \m}$, admits no consistent local
 deformation which (i) is Lorentz-invariant; and (ii) contains no
 more than two derivatives of the field [i.e., the allowed interaction
 terms under consideration contains at most $\partial^2 T$ or 
 $(\partial T)^2$].
 No restriction is imposed
 on the polynomial degree of the interaction. Our result confirms
 previous unsuccessful attempts \cite{Curt,Hull1,Hull20}.  We also demonstrate
 the rigidity, to first-order in the deformation parameter, of the
 algebra of the gauge symmetries without making any assumption on
 the number of derivatives.

 Besides their occurrence in dual formulations of linearized
 gravity, tensor fields in exotic representations of the Lorentz
 group arise in the long-standing related problem of constructing
 consistent interactions among particles with higher spins
 \cite{AB,Vasiliev,SezSun,Segal,Sagnotti}. A further motivation for the
 analysis of exotic higher spin gauge fields come from recent
 developments in $M$-theory, where a matching between
 the $D=11$ supergravity equations \cite{CJS} and the $E_{10 \vert
 +10}/K(E_{10})$ coset model equations ($K(E_{10})$ being the
 maximal compact subgroup of the split form of $E_{10\vert +10}$ of
 $E_{10}$) was exhibited up to height $30$ in the $E_{10}$ roots \cite{Damour}
 (the relevance of $E_{10}$ in the supergravity context was
 indicated much earlier in \cite{Julia}). One possibility for going
 beyond this height would be to introduce additional higher spin
 fields, most of which would be in exotic representations of the
 Lorentz group. Indeed, a quick argument shows that such fields
 might yield the exponentials associated with the higher height
 $E_{10}$-roots -- if they can be consistently coupled to gravity,
 an unsolved problem so far.  The introduction of such additional
 massless fields would also be in line with what one expects from
 string theory (in the high energy limit where the string tension
 goes to zero \cite{Gross}). 
 The same motivations come from the covariant coset construction of 
 \cite{West} where $D=11$ supergravity is conjectured to provide a 
 non-linear realization of $E_{11}$.
 The dual tensor field $T_{\l_1 \l_2
 \cdots \l_{8} \m}$ has actually already been identified in
 connection with both the $E_{11}$ \cite{West} and the $E_{10}$ 
 roots \cite{Damour,Pio}.  Note
 that mixed symmetry fields appear also in the models of
 \cite{BTZ,Burdik:2001hj}.

 In order to investigate the consistent, local, smooth deformations
 of the theory, we shall follow the cohomological approach of
 \cite{BH}, based on the antifield formalism \cite{BV,BVbis,Gomis}.
 An alternative, Hamiltonian based deformation point of view has
 been developed in \cite{Bizdadea}.  One advantage of the
 cohomological approach, besides its systematic aspect, is that it
 minimizes the work that must be done because most of the necessary
 computations are either already in the literature \cite{GK} or are
 direct extensions of existing developments carried out for
 $1$-forms \cite{BBH1,BBH2}, $p$-forms \cite{HK} or gravity
 \cite{BDGH,Beketal} (see also \cite{BrTh,Anco} for recent developments on
 the $1$-form-$p$-form case). To a large extent, our no-go theorem
 is obtained by putting together, in a standard fashion, various
 cohomological computations which have an interest in their own
 right and which have been already published or can be obtained
 through by-now routine techniques.

 \section{The free theory}
 \setcounter{equation}{0} \setcounter{theorem}{0}
 \setcounter{lemma}{0} \subsection{Lagrangian, gauge symmetries}

 As stated above, we first restrict the explicit analysis to the case of a tensor
 $T$ with $3$ indices, $T = T_{\a \b \m}$, which is dual to
 linearized gravity in $D=5$ (but we shall carry the analysis
 without specifying $D$, taken only to be stricly greater than $4$,
 $D>4$, so that the theory carries local degrees of freedom). The
 symmetry properties read \be T_{\a\b\g} = T_{[\a\b]\g}, \; \; \;
 T_{[\a\b]\g}+T_{[\b\g ]\a}+T_{[\g\a ]\b} = 0. \label{ident} \ee

 As shown in \cite{BB,dMH}, the appropriate algebro-differential
 language for discussing gauge theories involving exotic
 representations of the Lorentz group is that of multiforms, or
 more accurately, that of
 hyperforms\footnote{Hyperforms are in irreps of the general linear group,
 while multiforms sit in reducible ones.} \cite{Olver,BB}.
 Multiforms were discussed recently in \cite{DVH2} as an auxiliary
 tool for investigating questions concerning $N$-complexes
 associated with higher spin gauge theories. It turns out that
 hyperforms have been introduced much earlier in the mathematical
 literature by Olver in the analysis of higher order Pfaffian
 systems with integrability criteria (Olver, unpublished work \cite{Olver}).
 We shall not use here the language of multiforms or hyperforms, however,
 because the relevant tensors involve only a few indices.

 The Lagrangian for the gauge tensor field $T_{\l_1 \l_2\m}$ reads
 \be {\cal
 L}=-\frac{1}{12}\left(F_{[\a\b\g]\d}F^{[\a\b\g]\d}-3F_{[\a\b\x]}
 ^{\ \ \ \ \x}F_{\ \ \ \ \ \l}^{[\a\b\l]}\right)\,, \label{Lagrangian}
 \ee 
 where $F$ is the tensor \be  F_{[\a\b\g ]\d}=\pa_{\a} T_{[
 \b\g ]\d}+\pa_{\b}T_{[\g\a]\d} +\pa_{\g}T_{[\a\b]\d}\equiv
 3\pa_{[\a}T_{\b\g]\d} .\ee 
 The gauge invariances are 
 \be
 \delta_{\s,\a} T_{[\a\b]\g}=2(\pa_{[\a} \s_{\b]\g} +\pa_{[\a}
 \a_{\b]\g}-\pa_{\g}\a_{\a\b})\,,
 \label{invdejauge}\\
 \ee 
 where $\s_{\a \b}$ and $\a_{\a \b}$ are arbitrary symmetric
 and antisymmetric tensor fields.  The tensor $F$ is invariant
 under the $\s$-gauge symmetries, but not under the $\a$-ones.  To
 get a completely gauge-invariant object, one must take one
 additional derivative. The tensor  
 \be
 E_{[\a\b\d][\e\g]}\equiv\frac12
 (\partial_{\e}F_{[\a\b\d]\g}-\partial_{\g}F_{[\a\b\d]\e})
 \label{fieldstrength} \ee 
 is easily verified to be gauge
 invariant.
 Moreover its vanishing implies that $T_{[\a\b]\g}$ is pure-gauge \cite{BB}.
 The most general gauge invariant object depends on the
 field $T_{\a \b \m}$ and its derivatives only through the
 ``curvature" $E_{[\a\b\d][\e\g]}$ and its derivatives.  It is
 convenient to define the Ricci-like tensor $E_{[\a\b]\g}$ and its
 trace : \be E_{[\a\b]\g}= \h^{\e\d} E_{[\a\b\d][\e\g]}, \; \; \; \;
 E_{\a}=\h^{\b\g}E_{[\a\b]\g}.
 \ee
 The equations of motion are then 
 \be \frac{\delta {\cl}}{\d
 T_{[\a\b]\g}}=3[E^{[\a\b]\g}+\h^{\g[\a}E^{\b]}]=0\,. \label{ELeq}
 \ee 
 Because the action is gauge-invariant, the equations of motion
 fulfill the ``Bianchi identities" 
 \be
 \pa_{\a}(E^{[\a\b]\g}+\h^{\g[\a}E^{\b]})\equiv 0\,.
 \label{Bianchi}
 \ee 
 One easy way to check these identities is to
 observe that one has \be \frac{\d\cl}{\d T_{[\m\n]\r}}\equiv
 \pa_{\l}G^{\l\m\n\r} \ee where the tensor $G^{\l\m\n\r}$ is
 completely antisymmetric in its first three indices,
 $G^{\l\m\n\r}=G^{[\l\m\n]\r}$. Explicitly, \be
 G^{\l\m\n\r}=\frac{3}{2}\Big( \pa^{[\l}T^{\m\n]\r}-\h^{\r\l}
 \pa^{[\m}T^{\n\a]}_{~~~\a}-\h^{\r\m}\pa^{[\n}T^{\l\a]}_{~~~\a}
 -\h^{\r\n}\pa^{[\l}T^{\m\a]}_{~~~\a} \Big). \ee

 The gauge symmetries (\ref{invdejauge}) are reducible. Indeed, \be
 \d_{\tilde{\s},\,\tilde{\a}}T_{[\a\b]\g}\equiv 0 \ee when \be
 \tilde{\s}_{\a\b}=6\pa_{(\a}\g_{\b)},
 ~~~\tilde{\a}_{\a\b}=2\pa_{[\a}\g_{\b]} \label{reduc} \ee where
 $\g_{\a}$ are {\it{arbitrary fields}}. There is no further local
 reducibility identity.

 The problem of introducing (smooth) consistent interactions is
 that of smoothly deforming the Lagrangian (\ref{Lagrangian}), \be
 {\cal L} \rightarrow {\cal L} + g {\cal L}_1 + g^2 {\cal L}_2 +
 \cdots, \ee the gauge transformations (\ref{invdejauge}), \be
 \delta_{\s,\a} T_{[\a\b]\g} = \hbox{(\ref{invdejauge})} + g
 \,\delta_{\s,\a}^{(1)} T_{[\a\b]\g} + g^2 \, \delta_{\s,\a}^{(2)}
 T_{[\a\b]\g} + \cdots \ee  and the reducibility relations
 (\ref{reduc}) in such a way that (i) the new action is invariant
 under the new gauge symmetries; and (ii) the new gauges symmetries
 reduce to zero on-shell when the gauge parameters fulfill the new
 reducibility relations.  By developing these requirements order by
 order in the deformation parameter $g$, one gets an infinite number
 of consistency conditions, one at each order.

 We shall impose the further requirement that the first order
 vertex ${\cal L}_1$ be Lorentz-invariant.  Under this sole
 condition (together with consistency), we show that one can always
 redefine the fields and the gauge parameters in such a way that
 the gauge structure is unaffected by the deformation (to first
 order in $g$).  That is, the gauge transformations remain abelian
 and the reducibility relations remain unchanged (``rigidity of the
 gauge algebra"). We next restrict the deformations to contain at
 most two derivatives of the fields, as the original free
 Lagrangian. This still leave a priori an infinite number of
 possibilities, of the schematic form $T^k (\partial T)^2$ where
 $k$ is arbitrary (a term $T^l
 \partial ^2 T$ is of course equivalent to $T^{l-1} (\partial T)^2$
upon integration by parts). We show, however, that within this
infinite class, there is no non-trivial deformation.  Any
deformation can be redefined away by a local change of field
variables.

 \subsection{BRST differential}

 As shown in \cite{BH}, the first-order consistent local
 interactions correspond to elements of the cohomology $H^{D,0}(s
 \vert d)$ of the BRST differential $s$ modulo the spacetime
 exterior derivative $d$, in maximum form degree $D$ and in ghost
 number $0$. That is, one must compute the general solution of the
 cocycle condition \be s a + db =0, \label{coc}\ee where $a$ is a
 $D$-form of ghost number zero and $b$ a $(D-1)$-form of ghost
 number one, with the understanding that two solutions $a$ and $a'$
 of (\ref{coc}) that differ by a trivial solution \be a' = a + s m
 + dn \ee should be identified as they define the same interactions
 up to field redefinitions. The cochains $a$, $b$, {\it{etc}} that appear
 depend polynomially on the field variables (including ghosts and
 antifields) and their derivatives up to some finite order (``local
 polynomials).  Given a non trivial cocycle $a$ of $H^{D,0}(s \vert
 d)$, the corresponding first-order interaction vertex ${\cal L}_1$
 is obtained by setting the ghosts equal to zero.

 According to the general rules, the spectrum of fields and
 antifields is given by
 \begin{itemize}
 \item the fields $T_{[\a\b]\g}$, with ghost number zero and
 antifield number zero; \item the ghosts $S_{(\a\b)}$ and
 $A_{[\a\b]}$ with ghost number one and antifield number zero;
 \item the ghosts of ghosts $C_{\a}$ with ghost number two and
 antifield number zero, which appear because of the reducibility
 relations; \item the antifields $T^{* [\a\b]\g}$, with ghost
 number minus one and antifield number one; \item the antifields
 $S^{*(\a\b)}$ and $A^{*[\a\b]}$ : ghost number minus two and
 antifield number two; \item the antifields $C^{*\a}$ with ghost
 number three and antighost number three.
 \end{itemize}

 The antifield number is also called ``antighost number''.  Since the
 theory at hand is a free theory, the BRST differential takes the
 simple form \be s = \delta + \gamma \ee The decomposition of $s$
 into $\delta$ plus $\gamma$ is dictated by the antifield number :
 $\delta$ decreases the antifield number by one unit, while
 $\gamma$ leaves it unchanged. Combining this property with $s^2
 =0$, one concludes that \be \delta^2 = 0, \; \delta \gamma +
 \gamma \delta = 0, \; \gamma^2 = 0. \ee A grading is associated to
 each of these differentials : $\g$ increases by one unit the
 ``pure ghost number" denoted {\it{puregh}} while $\d$ increases
 the ``antighost number'' {\it{antigh}} by one unit. The ghost
 number {\it{gh}} is defined by \be
 {\it{gh}}={\it{puregh}}-{\it{antigh}}. \ee

 The action of the differentials $\gamma$ and $\delta$ on all the
 fields of the formalism is displayed in the following array which
 indicates also the  pureghost number, antighost number, ghost
 number and grassmannian parity of the various fields :
 \vspace{5mm}

 \begin{center}
 \begin{tabular}{|c|c|c|c|c|c|c|}
 \hline Z & $\gamma(Z)$  & $\delta(Z)$  & $puregh(Z)$  &
 $antigh(Z)$  & $gh(Z)$  & parity \\ \hline
 $T_{[\a\b]\g}$  &  $\g T_{[\a\b]\g}$  & $0$  &$0$  & $0$  &$0$ &$0$ \\
 $S_{(\a\b)}$ & $6\pa_{(\a}C_{\b)}$ & $0$ & $1$ & $0$ & $1$ & $1$ \\
 $A_{[\a\b]}$ & $2\pa_{[\a}C_{\b]}$ & $0$ & $1$ & $0$ & $1$ & $1$ \\
 $C_{\a}$       & $0$ &  $0$      & $2$  & $0$ & $2$ &  $0$ \\
 $T^{*[\a\b]\g}$ & $0$ & $3[E^{[\a\b]\g}+\h^{\g[\a}E^{\b]}]$ & $0$
 & $1$ & $-1$ & $1$ \\
 $S^{*\a\b}$ & $0$ & $-\pa_{\g}(T^{*[\g\a]\b}+T^{*[\g\b]\a})$
 & $0$ & $2$ & $-2$ & $0$
 \\
 $A^{*\a\b}$ & $0$ & $-3\pa_{\g}(T^{*[\g\a]\b}-T^{*[\g\b]\a})$
 & $0$ & $2$ & $-2$ & $0$ \\
 $C^{*\a}$ & $0$ & $6\pa_{\m}S^{*\m\a}+2\pa_{\m}A^{*\m\a}$ & $0$&
 $3$ & $-3$ & $1$
 \\
 \hline
 \end{tabular}
 \end{center}
 where $\g
 T_{[\a\b]\g}=2(\pa_{[\a}S_{\b]\g}+\pa_{[\a}A_{\b]\g}-\pa_{\g}A_{\a\b})$.
 \vspace{5mm}

 It is convenient to perform a change of variables in the
 $antigh=2$ sector in order for the Koszul-Tate differential to
 take a simpler expression when applied on all the antifields of
 {\it{antigh}} $\geq$ $2$. We define \be C^{*\a\b}=3S^{*\a\b} +
 A^{*\a\b}. \ee It leads to the following simple expressions \bqn
 \d C^{*\a\b}&=&-6 \pa_{\g} T^{*[\g\a]\b},
 \\
 \d C^{*\m}&=&2\pa_{\n}C^{*\n\m}. \eqn

 \subsection{Strategy}

 To compute $H^{D,0}(s \vert d)$, one proceeds as in
 \cite{BBH1,BBH2} : one expands the cocycle condition $sa + db = 0$
 according to the antifield number.  To analyse this resulting
 equations, one needs to know the cohomological groups
 $H(\g)$, $H(\g \vert d)$ in strictly positive antighost number,
 $H(\d \vert d)$ and $H^{inv}(\d \vert d)$.

 \section{Standard results}
 \setcounter{equation}{0} \setcounter{theorem}{0}
 \setcounter{lemma}{0} Of the cohomologies just listed, some are
 already known while some can be computed straightforwardly.
 \subsection{Cohomology of $\g$}
 \label{cohog} The cohomology of $\g$ (space of solutions of $\g a
 = 0$ modulo trivial coboundaries of the form $\g b$) has been
 explicitly worked out in \cite{GK} and turns out to be generated
 by the following variables,
 \begin{itemize}
 \item  the antifields and all their derivatives, denoted by
 $[\Phi^*]$, \item  the undifferentiated ghosts of ghosts
 $C_{\m}$,
 \item the following ``field strength" components of the  ghosts
 $A_{[\a\b]}$ : $H^A_{[\a\b\g]}\equiv\pa_{[\a}A_{\b\g]}$ (but not
 their derivatives, which are exact), \item the $T$-field strength
 components defined in (\ref{fieldstrength}) and all their
 derivatives denoted by $[E_{[\a\b\g][\d\e]}]$.
 \end{itemize}
 Therefore, the cohomology of $\g$ is isomorphic to the algebra \be
 \left\{ f\left([E_{[\a\b\g][\d\e]}],
 [\Phi^*],C_{\m},H^A_{[\a\b\g]}\right) \right\} \ee of functions of
 the generators.  The ghost-independent polynomials
 $\a([E_{[\a\b\g][\d\e]}],[\Phi^*])$ are called ``invariant
 polynomials".

 \vspace{.3cm} \noindent {\bf Comments} \vspace{.1cm}

 Let $\left\{\omega^I\left(C_{\m},H^A_{[\a\b\g]}\right)\right\}$ be
 a basis of the algebra of polynomials in the variables $C_{\m}$
 and $H^A_{[\a\b\g]}$. Any element of $H(\g)$ can be decomposed in
 this basis, hence for any $\g$-cocycle $\a$
 \be\gamma\a=0 \quad\Leftrightarrow\quad
 \a=\a_I([E_{[\a\b\g][\d\e]}],[\Phi^*])\;
 \omega^I\left(C_{\m},H^A_{[\a\b\g]}\right) + \g \b
 \label{gammaa}\ee where the $\a_I$ are invariant polynomials.
 Furthermore, $\a_I\omega^I$ is $\g$-exact if and only if all the
 coefficients $\a_I$ are zero \be \a_I\omega^I=\gamma\b,\quad
 \Leftrightarrow\quad \a_I=0,\quad\mbox{for
 all}\,\,I.\quad\label{gammab}\ee Another useful property of the
 $\o^I$ is that their derivatives are $\g$-exact and thus, in
 particular, \be d \o^I = \g \hat{\o}^I \ee for some $\hat{\o}^I$.

 \subsection{General properties of $H(\g \vert d)$}
 The cohomological space $H(\g \vert d)$ is the space of
 equivalence classes of forms $a$ such that $\g a+db=0$, identified
 by the relation $a\sim a'$ $\Leftrightarrow$ $a'=a+\g c+df$. We
 shall need properties of $H(\g \vert d)$ in strictly positive
 antighost (= antifield) number.  To that end, we first recall the
 following theorem on invariant polynomials (pure ghost number
 $=0$) :
 \begin{theorem}\label{2.2}
 In form degree less than n and in antifield number strictly
 greater than $0$, the cohomology of $d$ is trivial in the space of
 invariant polynomials.
 \end{theorem}
 The argument runs as in \cite{BBH1,BBH2}, to which we refer for
 the details.

 \vspace{.3cm}  Theorem \ref{2.2}, which deals with $d$-closed
 invariant polynomials that involve no ghosts (one considers only
 invariant polynomials), has the following useful consequence on
 general $\g$-mod-$d$-cocycles with $antigh
 >0$.

 \vspace{.3cm} \noindent {\bf{Consequence of Theorem \ref{2.2}}}
 \vspace{.1cm}

 {\it{If $a$ has strictly positive antifield number (and involves
 possibly the ghosts), the equation $\gamma a + d b = 0$ is
 equivalent, up to trivial redefinitions, to $\gamma a = 0.$ That
 is,
 \begin{equation}
 \left.
 \begin{array}{c}
 \gamma a + d b = 0, \\
 antigh(a)>0
 \end{array}
 \right\} \Leftrightarrow \left\{
 \begin{array}{c}
 \gamma a' = 0\,, \\
 a'=a+dc
 \end{array}\right.\,.\label{blip}
 \end{equation}
 Thus, in antighost number $>0$, one can always choose
 representatives of $H(\g \vert d)$ that are strictly annihilated
 by $\g$}}.  Again, see \cite{BBH1,BBH2}.

 \subsection{Characteristic cohomology $H(\d \vert d)$}
 We now turn to the groups $H(\d \vert d)$, i.e., to the solutions
 of the condition $\d a + db = 0$ modulo trivial solutions of the
 form $\d m + dn$.  As shown in \cite{BBH1}, these groups are
 isomorphic to the groups $H(d \vert \d)$ of the characteristic
 cohomology, describing ordinary and higher order conservation laws
 (i.e., $n$-forms built out of the fields and their derivatives
 that are closed on-shell). Without loss of generality, one can
 assume that the solution $a$ of $\d a + db = 0$ does not involve
 the ghosts, since any solution that vanishes when the ghosts are
 set equal to zero is trivial \cite{MH91}. By applications of the
 results and methods of \cite{BBH1}, one can establish the
 following theorems (in $H^D_q(\delta \vert d)$, $D$ is the form degree and
 $q$ the antighost (= antifield) number) :

 \begin{theorem}
 \label{vanishing} The cohomology groups $H^D_q(\delta \vert d)$
 vanish in antifield  number $q$ strictly greater than $3$, \be
 H^D_q(\delta \vert d) = 0 \, \hbox{ for } q>3. \ee
 \end{theorem}

 \begin{theorem}
 \label{conservation2} A complete set of representatives of
 $H^D_3(\d \vert d)$ is given by the antifields $C^{*\m}$ conjugate
 to the ghost of ghosts, {\it{i.e.}}, \be \d a^D_3+d a^{D-1}_{2}=0
 \Rightarrow a^D_3=\l_{\m}C^{*\m} dx^0\wedge dx^1\wedge\ldots\wedge
 dx^{D-1}+\d b_4^D+d b_3^{D-1} \ee where the $\l_{\m}$ are
 constants.
 \end{theorem}

 \begin{theorem}
 \label{conservation3} In antighost number $2$, the general
 solution of \be \d a^D_2+d a^{D-1}_{1}=0 \ee reads, modulo trivial
 terms, \be a^D_2 = C^{*\m \n} t_{\m \n \r} x^\r \, dx^0\wedge
 dx^1\wedge\ldots\wedge dx^{D-1}\ee where $t_{\m \n \r}$ is an
 arbitrary, completely antisymmetric, constant tensor, $t_{\m \n
 \r} = t_{[\m \n \r]}$.  If one considers cochains $a$ that have no
 explicit $x$-dependence (as it is necessary for constructing
 Poincar\'e-invariant Lagrangians), one thus find that the
 cohomological group $H^D_2(\d \vert d)$ vanishes.
 \end{theorem}

 \noindent {\bf Comment}

 The cycle $C^{*\m}$ is associated to the conservation law $d \; \!
 ^* G \approx 0$ for the $(D-3)$-form $\! ^* G $ dual to
 $G^{[\l\m\n]\r}$ (the equations of motion read $\partial_\l
 G^{\l\m\n\r} \approx 0$ ). The cycle $C^{*\m \n} t_{\m \n \r}
 x^\r$ is associated to the conservation law $\partial_\lambda I^{\lambda\sigma\m\n\r}\approx 0$
 where $I^{\lambda\sigma\m\n\r}$ is equal to the tensor $G^{\lambda\sigma\m\n}x^{\r}
 +3\eta^{\lambda\mu}T^{\n\r\sigma}-3\eta^{\lambda\mu}\eta^{\sigma\nu}T^{\a\r}_{\quad\a}$
 completely antisymmetrized in the three indices $\m$, $\n$, $\r$ and in the pair $\l$, $\sigma$.
 The above theorems provide a complete description of $H^D_k(\delta
 |n)$ for $k>1$ and show that these groups are finite-dimensional.
 In contrast, the group $H^{D}_1 (\delta |d)$, which is related to
 ordinary conserved currents, is infinite-dimensional since the
 theory is free.  It is not computed here.

 \subsection{Invariant characteristic cohomology : $H^{inv}(\d \vert
 d)$} The crucial result that underlies all consistent interactions
 deals not with the general cohomology of $\delta$ modulo $d$ but
 rather with the {\em invariant} cohomology of $\delta$ modulo $d$.
 The group $H^{inv}(\delta \vert d)$ is important because it
 controls the obstructions to removing the antifields from a
 $s$-cocycle modulo $d$, as we shall see explicitly below.

 The central theorem that gives $H^{inv}(\delta \vert d)$ in
 antighost number $\geq 2$ is
 \begin{theorem}\label{2.6}
 Assume that the invariant polynomial $a_{k}^{p}$
 ($p =$ form-degree, $k =$ antifield number)
 is $\delta$-trivial
 modulo $d$,
 \be
 a_{k}^{p} = \delta \mu_{k+1}^{p} + d \mu_{k}^{p-1} ~ ~ (k \geq 2).
 \label{2.37}
 \ee
 Then, one can always choose $\mu_{k+1}^{p}$ and $\mu_{k}^{p-1}$ to be
 invariant.
 \end{theorem}
 Hence, we have $H^{n,inv}_k(\delta \vert d) = 0$ for $k>3$ while
 $H^{n,inv}_3(\delta \vert d)$ is given by Theorem
 \ref{conservation2} and $H^{n,inv}_2(\delta \vert d)$ vanishes (in
 the space of translation-invariant cochains), by Theorem
 \ref{conservation3}.

 The proof of this theorem proceeds exactly as the proofs of
 similar theorems established for vector fields \cite{BBH2},
 $p$-forms \cite{HK} or gravity \cite{BDGH}.  We shall therefore
 skip it.

 \section{Rigidity of the gauge algebra}
 \setcounter{equation}{0} \setcounter{theorem}{0}
 \setcounter{lemma}{0} We can now proceed with the derivation of
 the cohomology of $s$ modulo $d$ in form degree $D$ and in ghost
 number zero. A cocycle of $H^{0,D}(s \vert d)$ must obey \be s a +
 d b = 0 \label{cocycsd} \ee  (besides being of form degree $D$
 and of ghost number $0$). To analyse (\ref{cocycsd}), we expand
 $a$ and $b$ according to the antifield number, $a = a_0 + a_1 +
 ...+a_k $, $b = b_0 + b_1 + ...+b_k  $, where, the expansion stops
 at some finite antifield number \cite{BBH2}. We recall \cite{BH}
 (i) that the antifield-independent piece $a_0$ is the deformation
 of the Lagrangian; (ii) that $a_1$, which is linear in the
 antifields $T^{*[\a\b]\g}$ contains the information about the
 deformation of the gauge transformations of the fields, given by
 the coefficients of $T^{*[\a\b]\g}$; (iii) that $a_2$ contains the
 information about the deformation of the gauge algebra (the term
 $C^*_Af^A_{~BC}C^BC^C$ with $C^*_A\equiv S^{*\a\b}, A^{*\a\b}$ and
 $C^A\equiv S_{\a\b},A_{\a\b}$ gives the deformation of the
 structure functions appearing in the commutator of two gauge
 transformations, while the term $T^*T^*CC$ gives the on-shell
 terms) and about the deformation of the reducibility functions
 (terms containing the ghosts of ghosts and the antifields
 conjugate to the ghosts); and (iv) that the $a_k$ ($k>3$) give the
 information about the deformation of the higher order structure
 functions, which appear only when the algebra does not close
 off-shell.  Thus, if one can show that the most general solution
 $a$ of (\ref{cocycsd}) stops at $a_1$, the gauge algebra is rigid :
 it does not get deformed to first order.

 Writing $s$ as the sum of $\gamma$ and $\delta$, the equation $s a
 + d b = 0$ is equivalent to the system of equations $\delta a_i +
 \gamma a_{i-1} + db_{i-1} = 0$ for $i = 1, \cdots, k$, and $\gamma
 a_k + db_k =0$.

 \subsection{Terms $a_k$, $k>3$}
 To begin with, let us assume $k > 3$. Then, using the consequence
 of theorem \ref{2.2}, one may redefine $a_k$ and $b_k$ so that
 $b_k = 0$, i.e., $\gamma a_k =0$. Then, $a_k = \alpha_J \omega^J$
 (up to trivial terms), where the $\alpha_J$ are invariant
 polynomials and where the $\{\omega^J\}$ form a basis of the
 algebra of polynomials in the variables $C_{\m}$ and
 $H^A_{[\a\b\g]}$. Acting with $\gamma$ on the second to last
 equation and using $\gamma ^2 =0$~, $\gamma a_k = 0$~, we get $d
 \gamma b_{k-1} = 0$~{\it{i.e.}} $\g b_{k-1}+dm_{k-1}=0 $; and
 then, thanks again to the consequence of theorem \ref{2.2},
 $b_{k-1}$ can also be assumed to be invariant, $b_{k-1} = \beta _J
 \omega ^J$. Substituting these expressions for $a_k$ and $b_{k-1}$
 in the second to last equation, we get : \be
 \delta [\alpha_J \omega^J] + d[\beta_J \omega^J] = \gamma (\ldots).
 \label{2.53}
 \ee
 This equation implies \be [\delta \alpha_J + d \b_J ]\omega^J =
 \gamma (\ldots) \label{2.54} \ee because the exterior derivative
 of a $\omega^J$ is equivalent to zero in $H(\g)$. Then, as
 dicussed at the end of subsection \ref{cohog}, this leads to : \be
 \d \a_J + d \b_J = 0~, ~~~\forall ~J. \label{delmodd} \ee If the
 antifield number of $\a_J$ is strictly greater than 3, the
 solution is trivial, thanks to our results on the cohomology of
 $\d$ modulo $d$ : $\a_J=\d \m_J+d\n_J$. Furthermore, theorem
 \ref{2.6} tells us that $\m_J$ and $\n_J$ can be chosen
 invariants. This is the crucial place where we need theorem
 \ref{2.6}. Thus $a_k= (\d
 \m_J+d\n_J)\omega^J=s(\m_J\o^J)+d(\n_J\o^J)\pm \n_Jd\o^J$. The
 last term $\n_Jd\o^J$ is equal to $\n_Js\hat{\o}^J$
 and differs from the
 $s$-exact term $s(\pm \n_J\hat{\o}^J)$ by the term $\pm \d\n_J\hat{\o}^J$,
 which is of lowest antifield number. Trivial redefinitions enable
 one to set $a_k$ to zero. Once this is done, $\b_J$ must satisfy
 $d\b_J=0$ and is then exact in the space of invariant polynomials,
 $\b_J=d\r_I$, and $b_{k-1}$ can be removed by appropriate trivial
 redefinitions. One can next repeat the argument for  antifield
 number $k-1$, etc, until one reaches antifield number $3$. This
 case deserves more attention, but what we can stress already now
 is that {\it{we can assume that the expansion of}} $a$
 {\it{in}} $sa+db=0$
 {\it{stops at antifield number $3$ and takes the form}}
 $a = a_0 +a_1 + a_2+ a_3$  {\it{with}} $b = b_0 + b_1 + b_2$. Note
 that this result is independent of any condition on the number of
 derivatives or of Lorentz invariance.  These requirements have not
 been used so far. The crucial ingredient of the proof is that the
 cohomological groups $H^{inv}_k(\delta \vert d)$, which control
 the obstructions to remove $a_k$ from $a$, vanish for $k>3$.

 \subsection{Computation of $a_3$}
 We have now the following descent:
 \bqn
 \d a_1 + \g a_0 +d b_0 &=&0~,
 \label{5.4}
 \\
 \d a_2 + \g a_1 +d b_1 &=&0~,
 \label{5.5}
 \\
 \d a_3 + \g a_2 +d b_2 &=&0~,
 \label{5.6}
 \\
 \g a_3&=&0~. \eqn We write $a_3=\a_I\o^I$ and $b_2=\b_I\o^I$.
 Proceeding as before we find that a necessary (but not sufficient)
 condition for $a_3$ to be a non-trivial solution of (\ref{5.6}),
 so that $a_2$ exists, is that $\a_I$ be a non-trivial element of
 $H^n_3(\d\vert d)$. The Theorem \ref{conservation2} imposes then
 $\a_I \sim C^{*\m}$. We then have to complete this $\a_I$ with an
 $\o^I$ of ghost number $3$ in order to build a candidate
 $\a_I\o^I$ for $a_3$. There are a priori a lot of possibilities to
 achieve this, but if one demands Lorentz invariance, only two
 possibilities emerge: \bqn
 a_3&=&C^{*\m}H_{\m\a\b}H^{\a\n\r}H^{\b}_{~\n\r}~,
 \label{firsta_3}\\
 a_3&=&C^{*\m}\varepsilon_{\m\n\r\l\s}H^{\n\r\l}C^{\s}~,
 \label{seconda_3} \eqn where we recall that
 $H_{\m\a\b}\equiv\pa_{[\m}A_{\a\b]}$ $\in H(\g)$ at ghost number
 one, and $C^{\s}$ $\in H(\g)$ at ghost number two. Since $C^{*\m}$
 has antighost number three [{\it{i.e.}} ghost number $(-)3$], we
 indeed have two ghost-number-zero $a_3$ candidates. The first is
 quartic in the fields and, if consistent, would lead to a quartic
 interaction vertex.  The second is cubic and, if consistent, would
 lift to an $a_0$ which breaks $PT$ invariance.

 However, neither of these candidates can be lifted all the way to
 $a_0$.  Both get obstructed at antighost number one: $a_2$ exists,
 but there is no $a_1$ that solves (\ref{5.5}) (given $a_3$ and the
 corresponding $a_2$).  The computation is direct but not very
 illuminating and so will not be reproduced here.

 \subsection{Computation of $a_2$}
 Continuing with our analysis, we set $a_3=0$ and get the system
 \bqn \d a_1 + \g a_0 +d b_0 &=&0~, \label{5.7}
 \\
 \d a_2 + \g a_1 +d b_1 &=&0~,
 \label{5.8}
 \\
 \g a_2&=&0~. \eqn Now $a_2$ has to be found in $H^n_2(\d\vert d)$,
 but this latter group vanishes (in the space of
 translation-invariant deformations), as shown in theorem
 \ref{conservation3}.   We can thus conclude that there is no
 possibility of deforming the free theory to obtain an interacting
 theory whose gauge algebra is non-abelian. To obtain this no-go
 result we just asked for locality, Lorentz invariance and the
 assumption that the deformed theory reduces smoothly to the free
 one as the deformation parameter goes to zero.

 \section{No-go theorem}
 \setcounter{equation}{0} \setcounter{theorem}{0}
 \setcounter{lemma}{0}

 With $a_i = 0$ for $i>1$, the cocycle condition (\ref{cocycsd})
 reduces to \bqn \d a_1 + \g a_0 + db_0&=&0\,,
 \label{5.9} \\
 \g a_1&=&0\,. \label{5.10} \eqn The last equation forces $a_1$ to
 take the schematic form $a_1 = T^{*} H p(E,
 \partial E, ...)$ where the constant term in $p$ is zero because
 $T^{*[\a\b]\g}H_{\a\b\g}$ (the only $E$-independent possibility
 allowed by Lorentz invariance) is identically zero due to opposite
 symmetries for $T^{*[\a\b]\g}$ and $H_{\a\b\g}$. But all these
 candidates $a_1$ involve at least four derivatives (two in $E$,
 one in $H$, and one in $T^{*}$ -- which counts for one because $\d
 T^{*}$ contains two derivatives, see e.g. \cite{BBH2}), so we
 reject this possibility on the assumption that the interaction
 terms in the Lagrangian should not have more than two derivatives.
 Thus, $a_1 = 0$ and the deformations not only do not modify the
 gauge algebra, but actually also leave unchanged the gauge
 transformations of the field $T_{[\a\b]\g}$.

 We are then reduced to look for  $a_0$ solutions of $\g a_0 +d
 b_0=0$, i.e., for deformations of the Lagrangian which must be
 gauge invariant up to a total derivative.  Because these
 deformations are gauge invariant up to a total derivative, their
 Euler-Lagrange derivatives are strictly gauge invariant.  These
 Euler-Lagrange derivatives contains two derivatives of the fields
 and satisfies Bianchi identities of the type (\ref{Bianchi})
 (because of the gauge invariance of $\int a_0$).  It is easy to
 see that the only gauge-invariant object satisfying these
 conditions are the Euler-Lagrange derivatives of the original
 Lagrangian itself, so we conclude that $a_0 \sim {\cal
 L}$ : the deformation only changes the coefficient of the free
 Lagrangian and is not essential.  
 In fact, allowing for an $a_0$ containing three derivatives would not 
 change the conclusions. Indeed, starting from the first derivative of
 the curvature, $\partial E$, there is no way to contract the indices 
 in order to form a candidate $X^{[\a\b]\g}$ with the symmetries of the field
 $T^{[\a\b]\g}$. Hence, acceptable $a_0$ (other than the original Lagrangian)
 should involve at least four derivatives.
 This completes the proof of the
 rigidity of the free theory.
 \\
 \section{Comments and conclusions}
 \setcounter{equation}{0} \setcounter{theorem}{0}
 \setcounter{lemma}{0}

 We can summarize our results as follows :  under the hypothesis of
 \begin{itemize} \item locality and \item Lorentz invariance
 \end{itemize}
 there is no smooth deformation of the free theory which modifies
 the gauge algebra. If one further excludes deformations involving
 four derivatives or more in the Lagrangian, then there are just
 no smooth deformation of the free theory at all.

 Without this extra condition on the derivative order, one can
 introduce Born-Infeld-like interactions that involve powers of the
 gauge-invariant curvatures $E_{[\a \b \g] [\l \m]}$.  Such
 deformations modify neither the gauge algebra nor the gauge
 transformations.

 The same no-go result can easily be extended to a collection of
 fields $T_{[\a \b]\g}$, as in \cite{BDGH}, or to a system of one
 $T_{[\a \b]\g}$ and one Pauli-Fierz field $h_{\m \n}$.

We have considered here explicitly the dual formulation of gravity
in $D=5$ dimensions, with a $3$-index tensor $T_{\a \b \g}$.  The
general case of exotic representations $T_{\l_1 \l_2 \cdots
\l_{D-3} \m}$ with more indices in the first row, which is
relevant in $D$ spacetime dimensions, is dealt with similarly,
provided one takes into account the additional reducibility
identities that appear then.  There are more candidates analogous
to $a_3$ (but now in higher antighost number) but we have checked
that all these candidates are eliminated if one restricts the
derivative order to two (in fact, the higher derivative terms are
in any case probably obstructed, but we have not verified this
explicitly, the derivative argument being sufficient to rule them
out).  In the end (going though all $a_i$'s), one finds again that
there is no deformation with no more than two derivatives.

 {}Finally, we note that the same techniques can be used to analyse
 consistent deformations of more general exotic gauge fields.  It
 is planned to return to this question in the future.

 \section*{Acknowledgments}
 We are grateful to Peter Olver for sending us a copy of his
 unpublished paper \cite{Olver}. The work of N.B. and M.H. is
 supported in part by the ``Actions de Recherche Concert{\'e}es'' of
 the ``Direction de la Recherche Scientifique - Communaut{\'e}
 Fran{\c c}aise de Belgique'', by a ``P\^ole d'Attraction
 Interuniversitaire'' (Belgium), by IISN-Belgium (convention
 4.4505.86), by Proyectos FONDECYT 1970151 and 7960001 (Chile) and
 by the European Commission RTN programme HPRN-CT-00131, in which
 they are associated to K. U. Leuven.  The work of X.B is supported
 in part by the European Commission RTN programme HPRN-CT-00131.

 \end{document}